\documentclass[
aps,
twocolumn,
superscriptaddress,
showpacs,
amsmath,amssymb,
prl
]{revtex4-1}

\usepackage{graphicx}
\usepackage{dcolumn}
\usepackage{bm}
\usepackage{color}

\begin{document}

\title{Space-time crystals of trapped ions}

\author{Tongcang Li}
 \affiliation{NSF Nanoscale Science and Engineering Center, 3112 Etcheverry Hall, University of California, Berkeley, California 94720, USA}
\author{Zhe-Xuan Gong}
 \affiliation{Department of Physics, University of Michigan, Ann Arbor, Michigan 48109, USA}
 \affiliation{Center for Quantum Information, Institute for Interdisciplinary Information Sciences, Tsinghua University, Beijing, 100084, P. R. China}
\author{Zhang-Qi Yin}
 \affiliation{Center for Quantum Information, Institute for Interdisciplinary Information Sciences, Tsinghua University, Beijing, 100084, P. R. China}
 \affiliation{Key Laboratory of Quantum Information, University of Science and Technology of China, Chinese Academy of Sciences, Hefei, 230026, P. R. China}
 \author{H. T. Quan}
 \affiliation{Department of Chemistry and Biochemistry, University of Maryland, College Park, Maryland 20742, USA}
\author{Xiaobo Yin}
 \affiliation{NSF Nanoscale Science and Engineering Center, 3112 Etcheverry Hall, University of California, Berkeley, California 94720, USA}
 \author{Peng Zhang}
 \affiliation{NSF Nanoscale Science and Engineering Center, 3112 Etcheverry Hall, University of California, Berkeley, California 94720, USA}
 \author{L.-M. Duan}
 \affiliation{Department of Physics, University of Michigan, Ann Arbor, Michigan 48109, USA}
 \affiliation{Center for Quantum Information, Institute for Interdisciplinary Information Sciences, Tsinghua University, Beijing, 100084, P. R. China}
 \author{Xiang Zhang}
 \email{Corresponding author: xiang@berkeley.edu}
 \affiliation{NSF Nanoscale Science and Engineering Center, 3112 Etcheverry Hall, University of California, Berkeley, California 94720, USA}
\affiliation{Materials Sciences Division, Lawrence Berkeley National Laboratory, 1 Cyclotron Road, Berkeley, California 94720, USA}

\date{\today}

\begin{abstract}
 Spontaneous symmetry breaking can lead to the formation of  time crystals, as well as spatial crystals.   Here we propose a space-time crystal of trapped ions and a method to realize it experimentally by confining ions in a ring-shaped trapping potential with a static magnetic field. The ions spontaneously form a spatial ring crystal due to Coulomb repulsion. This ion crystal can rotate persistently at the lowest quantum energy state in magnetic fields with fractional fluxes. The persistent rotation of trapped ions produces the temporal order, leading to the formation of a space-time crystal. We show that these space-time crystals are robust for direct experimental observation. We also study the effects of finite temperatures on the persistent rotation. The proposed space-time crystals of trapped ions provide a new dimension for exploring many-body physics and emerging properties of matter.

\end{abstract}
\maketitle

Symmetry breaking plays profound roles in many-body physics and particle physics \cite{strocchi2008,Higgs1964}. The spontaneous breaking of continuous spatial translation symmetry to discrete spatial translation symmetry leads to the formation of various crystals in our everyday life. Similarly, the spontaneously breaking of time translational symmetry can lead to the formation of a time crystal \cite{wilczek2012,shapere2012}.  Due to  symmetry breaking, the effective ground state  of a system can be an inhomogeneous many-body state whose probability amplitude changes periodically in time \cite{wilczek2012}, having a temporal symmetry different from that of the true quantum ground state of the Hamiltonian. This is similar to the symmetry breaking during the formation of a spatial crystal, where the crystal has a symmetry
different from that of the underlying system \cite{Anderson1997}.

Intuitively, if a spatially ordered system rotates persistently in the lowest energy state, the system will reproduce itself periodically in time, forming a time crystal in analog of an ordinary crystal. Such a system looks like a perpetual motion machine and may seem implausible in the first glance. On the other hand, it has been known that a superconductor \cite{Byers1961,Zhu2010} or even a normal metal ring  \cite{BUTTIKER1983,Levy1990,Bleszynski-Jayich2009} can support persistent currents in its quantum ground state under proper conditions. However, the rotating Cooper pairs or electrons in a metal  are not quantum time crystals since their wavefunctions are homogeneous  and no time translational symmetry is broken. A soliton model that assumes strong attractive interaction between particles with the same type of charge has been proposed \cite{wilczek2012}, which is difficult to be realized experimentally. While it has been proved mathematically that time crystals can exist in principle \cite{wilczek2012,shapere2012}, it was not clear how to realize and observe time crystals experimentally.

In this paper, we propose a method to create a space-time crystal (Fig. 1(a)), which is also a time crystal, with cold ions in a cylindrically symmetric trapping potential. Different from electrons in conventional materials, ions trapped in vacuum have strong Coulomb repulsion between each other and have internal atomic states. The strong Coulomb repulsion between ions enables the spontaneous breaking of the spatial translation symmetry, resulting in the formation of a spatial order that can be mapped into temporal order. The internal atomic states of ions can be utilized to cool the ions to the ground state as well as observing their persistent rotation directly by state-dependent fluorescence.
 Different from Wilczek's soliton model \cite{wilczek2012},
our trapped-ion model is a crystal (in space) even when the magnetic flux is zero.

\begin{figure}[btp]
\setlength{\unitlength}{1cm}
\begin{picture}(8,11)
\put(0.5,0){\includegraphics[totalheight=11cm]{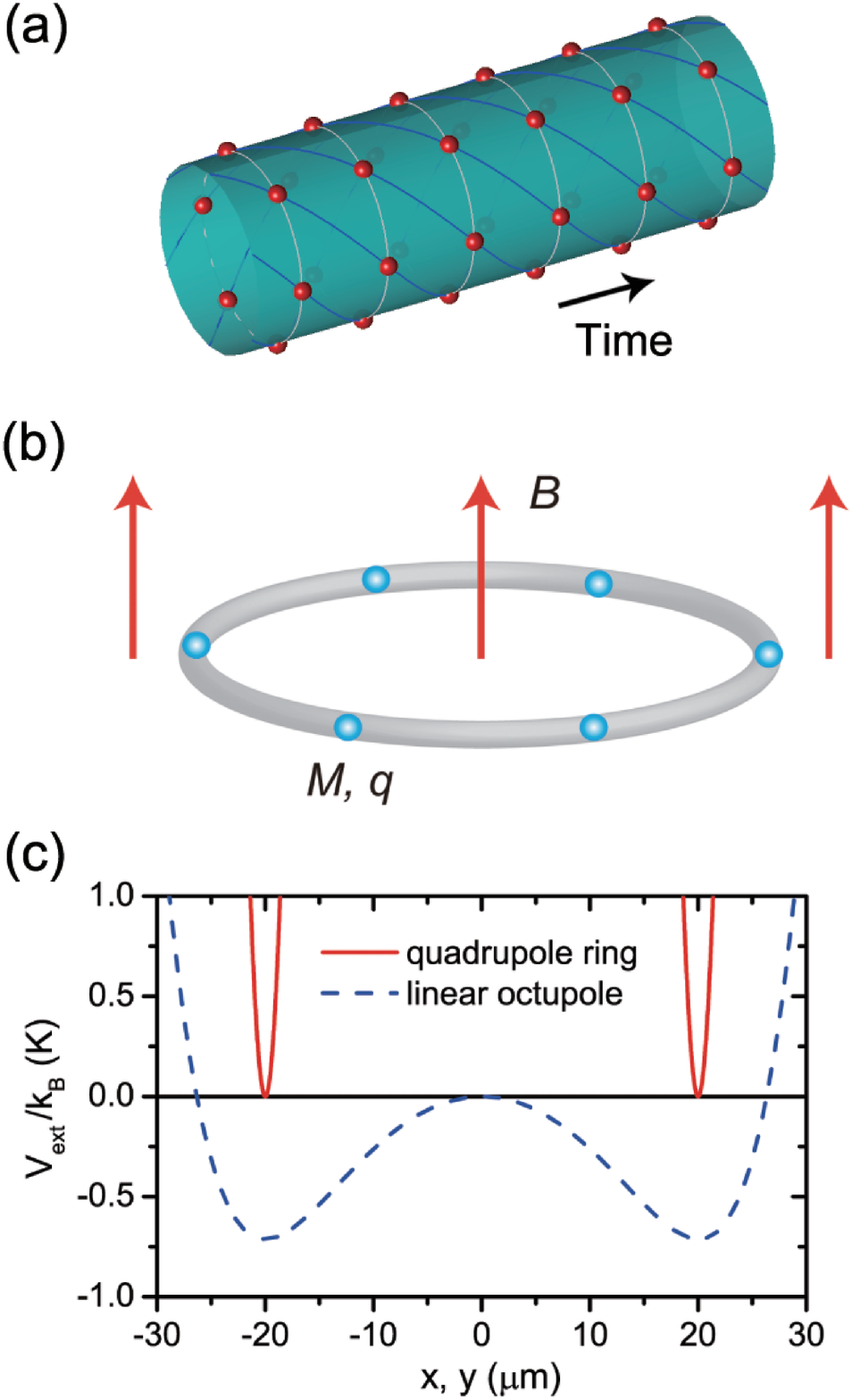}}
\end{picture}
\caption{Schematic of creating a space-time crystal. (a), A possible structure of a space-time crystal. It has periodic structures in both space and time. The particles rotate in one direction even at the lowest energy state. (b), Ultracold ions confined in a ring-shaped trapping potential in a weak magnetic field. The mass and charge of each ion are $M$ and $q$, respectively.  The diameter of the ion ring is $d$, and the magnetic field is $B$. (c), Examples of the pseudo-potentials ($V_{ext}$) for a $^9$Be$^+$ ion in a quadrupole ring trap (solid curve) and a linear octupole trap (dashed curve) along the $x$ or $y$ axis. }
\label{fig:scheme}
\end{figure}

Trapped ion Coulomb crystals have provided unique opportunities for studying many-body phase transitions \cite{Gong2010,Kim2010,Blatt2012} and quantum information science \cite{Cirac1995,Leibfried2003}. The two most common types of ion traps are the Penning trap \cite{Brown1986} and the Paul trap \cite{Leibfried2003}. The magnetron motion of an ion in the Penning trap is an orbit around the top of a potential hill that is not suitable for studying the behavior of the ion in its rotational ground state.  Here we propose to use a combination of a ring-shaped trapping potential from a variation of the Paul trap, and a weak static magnetic field. The ring-shaped trapping potential can be created by a quadrupole storage ring trap \cite{Schatz2001,Madsen2010}, a linear rf multipole trap \cite{Okada2007,Champenois2010}, a multiple trap from planar ring electrodes \cite{clark2012} or other methods. As an example, we consider $N$ identical ions of mass $M$ and charge $q$ in a ring trap and a uniform magnetic field $B$ (Fig. 1(b)). The magnetic field is parallel to the axis of the trap. It is very weak so that it does not affect trapping. The equilibrium diameter of the ion ring is $d$. Figure 1(c) shows examples of the trapping potentials for a $^9$Be$^+$ ion in the radial plane of a quadrupole ring trap and a linear octupole trap. See the supplemental material \cite{supplemental} for more information.

When the average kinetic energy of ions ($k_B T/2$, where $k_B$ is the Boltzmann constant and $T$ is the temperature) is much smaller than the typical Coulomb potential energy between ions, i.e. $T \ll Nq^2 /(2 \pi^2\epsilon_0 k_B d)$,  the ions form a Wigner ring crystal.  For ions in a ring crystal, we can expand the Coulomb potential around equilibrium positions to the second order. So the many-body Hamiltonian \cite{supplemental} becomes quadratic. We can diagonalize the quadratic Hamiltonian by introducing a set of $N$ normal coordinates $q_j$ and normal momenta $p'_j$.  The normal coordinate and momentum of the collective rotation mode are $q_1=\frac{1}{\sqrt{N}}\sum_j \theta_j$  and   $p'_1=\frac{1}{\sqrt{N}}\sum_j p_j$, respectively. The remaining $N-1$ normal coordinates correspond to relative vibration modes. Choosing the potential energy at equilibrium positions as the origin of energy, the Hamiltonian of the system becomes \cite{Burmeister2002}:
\begin{equation}
 \begin{aligned}
 \label{eq:rotHam2}
 H =\frac{2\hbar^2}{Md^2}[(-i\frac{\partial}{\partial q_1}-\sqrt{N} \alpha)^2+\sum^N_{j=2}(-\frac{\partial^2}{\partial q^2_j}+\eta^2\omega_j^2q^2_j)],
 \end{aligned}
\end{equation}
where $\hbar=h/(2\pi)$ and $h$ is the Planck constant,  $\alpha = q\pi d^2 B/(4h)$ is the normalized magnetic flux, $\eta^2=q^2Md/(8\pi \hbar^2 \epsilon_0)$, and $\omega_j$ ($j\geq 2$) is the normalized normal mode frequency.

The lowest normalized relative vibration frequency is $\omega_2 = 2.48$ when $N=10$ and will increases as $N$ increases \cite{Burmeister2002}. $\omega_2 \approx \sqrt {0.32 N \ln (0.77 N)}$ for large $N$.  $\eta=5.6\times 10^4$ for $^9\texttt{Be}^+$ ions in a $d=20 \,\mu$m ring trap. So it costs a lot of energy to excite the relative vibration modes. Thus we have $n_j=0$ for all $j\geq 2$ at lowest energy states, where $n_j$ are the occupation numbers of the relative vibration modes. The wavefunction  has to be symmetric with respect to the exchange of two identical bosonic ions (e.g., $^9\texttt{Be}^+$ ions), and has to be antisymmetric with respect to the exchange of two identical fermionic ions (e.g., $^{24}\texttt{Mg}^+$ ions).
For an ion ring of identical bosonic ions, the energy $E_{n_1}$ and the angular frequency $\omega_{n_1}$  of the $n_1$-th eigenstate of the collective rotation mode are \cite{supplemental}:
\begin{equation}
 \begin{aligned}
 \label{eq:results}
 E_{n_1} &=E^*(n_1-\alpha)^2=\frac{2N \hbar^2}{Md^2} (n_1-\alpha)^2,
 \\
  \omega_{n_1} &=\omega^*(n_1-\alpha)=\frac{4\hbar}{Md^2} (n_1-\alpha),
 \end{aligned}
\end{equation}
where $E^* = 2N \hbar^2/(Md^2)$ and $\omega^* = 4\hbar/(Md^2)$ are the characteristic energy and the characteristic frequency of the collective rotation, respectively.
For identical fermionic ions, the results are the same as Eq. (\ref{eq:results}) if $N$ is an odd number, and $n_1$ should be changed to $n_1+\frac{1}{2}$ if $N$ is an even number.

\begin{figure}[bh]
\centering
\includegraphics[width=8cm]{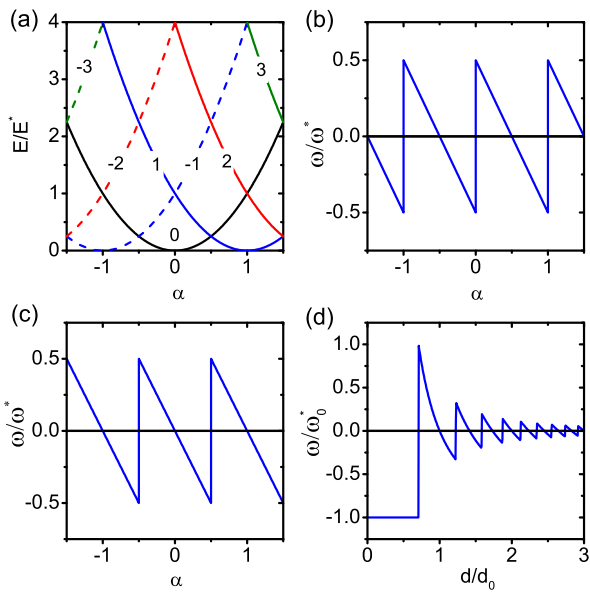}
\caption{ The energy levels and rotation frequencies of trapped ions. (a), The energy levels of   identical bosonic ions  as a function of the magnetic flux $\alpha$. The quantum number $n_1$ is labeled on each curve.  The angular frequency of the persistent rotation as a function $\alpha$ is shown in (b) for an even number of fermionic ions, and (c) for bosonic ions. (a) and (c) are also applicable to an odd number of fermionic ions. (d), The  angular frequency of the persistent rotation of a bosonic ion ring  in a constant magnetic field $B_0 > 0$  as a function of its normalized diameter $d/d_0$.}
\label{fig:energyvelocity}
\end{figure}

In classical mechanics, the angular velocity of the lowest energy state is always $\omega/\omega^* = 0$, which means that the ions do not rotate. In quantum mechanics, however, $\omega/\omega^* = 0$ is not an eigenvalue when the normalized magnetic flux  $\alpha$ is not an integer or half of an integer. So the ions can rotate persistently even at the ground state. Since the ions are in the ground state already, there is no radiation loss due to the rotation. The rotation frequency is independent of the number of ions in the ring. The energy gap between the ground state and the first excited state is $\Delta E=N \hbar^2/(Md^2)$ when $\alpha=1/4$. $\Delta E \rightarrow \infty$ when $N \rightarrow \infty$. Thus the persistent rotation of identical ions is a macroscopic quantum phenomenon and is robust for observation when $N$ is large, which is important for a time crystal \cite{wilczek2012}.  If the relative vibration modes are not in their ground states, the symmetry requirement of the center of mass motion of identical particles is relaxed and the energy gap between different center of mass motion states becomes smaller. The result becomes the same as that of a rigid body when all particles are different from each other. Then the maximum energy gap between the ground state and the first excited state is  $\Delta E_{rigid}=2 \hbar^2/(NMd^2)$.

Figure 2(a) and 2(c) show the lowest energy levels and rotation frequencies of an ion ring consisting identical bosonic ions. The angular frequency of the persistent rotation of the ground state is a periodic function of the magnetic flux (Fig. 2(c)). Figure 2(d) shows  the rotation frequency of a bosnic ion ring in a constant positive magnetic field $B_0$ as a function of its normalized diameter $d/d_0$, where $d_0=\sqrt{4 h/(\pi qB_0)}$. The rotation frequency in Fig. 2(d) is normalized by $\omega^*_0=q B_0/2M$. The ground state is $n_1=0$ when $d/d_0<1/\sqrt{2}$. The rotation frequency is independent of the ring diameter and the number of ions in the ring for this state, and oscillates and decreases to 0 when  $d/d_0$ increases above $1/\sqrt{2}$. If we confine many ions in a harmonic trap to form a 3D spatial crystal, ions in the crystal will rotate with the same angular frequency $\omega^*_0$ and form a 4D space-time crystal when the outer diameter of the ion crystal is smaller than $d_0/\sqrt{2}$. If we confine ions in two concentric ring traps with diameters larger than $d_0/\sqrt{2}$, the rotation frequencies of the two rings can be different or the same, depending on the interaction between ions in different rings. When the ratio of the rotation frequencies of the two rings is an irrational number, the ions have an order in time but cannot reproduce their positions simultaneously. Thus we have a time quasicrystal, in analog of a conventional spatial quasicrystal \cite{Levine1986}.

The persistent rotation of trapped ions can be detected by measuring the Doppler shift of moving ions, or inferred by probing the energy levels of the ion ring. More importantly, we can observe the persistent rotation directly by measuring the ion positions twice when $N$ is large. For example, if we have an ion ring consisting $N$ identical $^9$Be$^+$ ions in their lowest energy state, we can first use a pulse of two co-propagating laser beams to change the hyperfine state of one (or a small fraction) of the ions by stimulated Raman transition and use this ion as a mark (qubit memory coherence time greater than 10 s has been demonstrated with $^9$Be$^+$ ions \cite{Langer2005}). Both laser beams are parallel to the axis of the ion ring and have waists of $w_0$. We assume that the pulse is very weak so that on average less than one ion is marked. This two-photon process localized the position of the mark ion with an uncertainty of about $\Delta x \sim w_0/\sqrt{2}$.  The amplitude of the transverse momentum of each photon in a Gaussian beam with waist of $w_0$ is about $\hbar/w_0$. Thus the momentum of the ion ring is changed by about $\Delta p \approx \sqrt{2} \hbar/w_0$.  $\Delta p$ should be  smaller than the absolute value of the initial momentum of the ion ring, which is  $N \hbar/(2d)$ when $\alpha$   = 1/4. Thus the waists of lasers need to satisfy  $2\sqrt{2}d/N < w_0 < \sqrt{2}d$ in order to localize the position of an ion without significantly alter the initial momentum of the ion ring. This condition can be fulfilled when $N$ is large. Then we can use a global probe laser which is only scattered by the mark ion \cite{Myerson2008} (state-dependent fluorescence) to measure its angular displacement ($\Delta \theta_{mk}$) after a time separation $\Delta t$. The displacement of the mark is about $\Delta \theta_{mk} \approx \omega^*(n_1-\alpha) \Delta t$  when $N$ is large. The mark ion repeats its position when $\Delta t \approx 2 \pi l/[\omega^*(n_1-\alpha)]$ , where $l$ is an integer. After the measurement, we can cool the ions back to the ground state and repeat the experiment again.

 The process of observing the rotation of ions in the ground state also elucidates the concept of quantum time crystals \cite{wilczek2012}. A quantum time crystal requires the probability amplitude of its effective ground state to  change periodically in time, which means it should be a rotating inhomogeneous state. The rotating condition is satisfied by the none-zero angular momentum.
However, for both the soliton model \cite{wilczek2012} and our ion crystal model, the probability amplitude of the true ground state of the center of mass is homogeneous,  due to the cylindrical symmetry of the system.  The required spatial inhomogeneity comes from symmetry breaking, which can happen spontaneously when $N$ is infinite, or triggered by an observation (by the environment or an observer) when $N$ is finite. The spatial symmetry breaking can localize the center of mass  to a single position \cite{Anderson1997} and makes the effective ground state inhomogeneous \cite{wilczek2012}.  The resulted rotating inhomogeneous state is not the true ground state, but an effective ground state with its energy infinitely close to that of the true ground state when $N \rightarrow \infty$ \cite{Anderson1997}. For both the
soliton model and our ion crystal model, a weak observation of a single particle can localize
the center of mass of all particles and project the true ground state to a rotating effective ground state. In contrast, Cooper pairs
or electrons in a metal ring have no spatial order, so an observation of a
single particle can not localize the center of mass of all particles.

To study the effects of finite temperatures on persistent rotation, we assume  the trapped ions are at thermal equilibrium with temperature $T$.  Then the average angular frequency of the ions is
\begin{equation}
 \begin{aligned}
 \label{eq:averageomega}
 \overline{\omega} = \sum^{\infty}_{n_1=-\infty} \frac{\omega_{n_1}}{Z} e^{-E_{n_1}/k_B T} ,
 \end{aligned}
\end{equation}
where the partition function is $Z = \sum_{n_1}  e^{-E_{n_1}/k_B T}$.
It is convenient to define $T^* \equiv E^*/k_B = N\hbar \omega^*/2k_B$ as the characteristic temperature for the ring of ions. $T^*$ can  be considered as the phase transition temperature of the space-time crystal. It increases when $N$ increases. Figure 3(a) shows the average rotation frequency of an ion ring consisting identical bosonic ions as a function of the temperature.
 $T^*$ and $\omega^*$ as a function of the diameter of an ion ring (or an electron ring) are displayed in Fig.~3(b).  For a  $d=100\, \mu$m ion ring consisting 100  $^9 \texttt{Be}^+$ ions, $T^*=1.1$ nK and  $\omega^*=2.8$ rad/s. $T^*$  is larger for smaller ion rings. The characteristic temperature of an electron ring is much higher because of smaller mass of electrons.
 
 \begin{figure}[th]
\centering
\includegraphics[width=8cm]{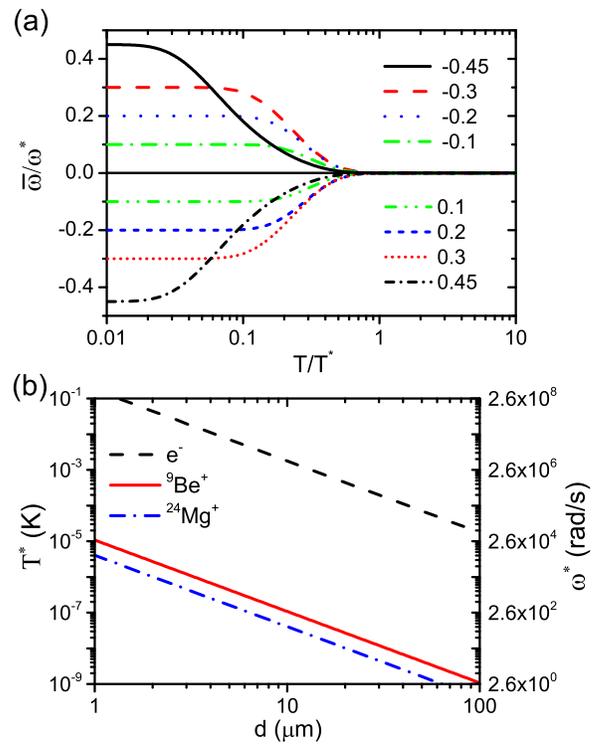}
\caption{ The temperature dependence of the persistent rotation of trapped ions. (a), The average angular frequency of the persistent rotation of identical bosonic ions as a function of the temperature. From top down, the magnetic flux   increases from -0.45 to 0.45. (b), The characteristic temperature (left axis) and the characteristic frequency (right axis) of the  persistent rotation of an ion (or electron) ring consisting 100 identical ions (or electrons) as a function of the diameter. From top down, the trapped particles are electrons, $^9$Be$^+$ ions, and $^{24}$Mg$^+$ ions.}
\label{fig:temperature}
\end{figure}

In order to experimentally realize such a space-time crystal with trapped ions, we need to confine ions tightly to have a small $d$, and cool the ions to a very low temperature. Recently, cylindrical ion traps with inner radius as small as 1 $\mu$m have been fabricated \cite{Cruz2007}. Simulations suggested that it is also possible to confine charged particles with a nanoscale rf trap \cite{Segal2006}. The challenge is that ions must be cooled to below 1 $\mu$K for a microscale trap (Fig. 3(b)). We propose to first add a pinning potential to confine ions with MHz trapping frequencies in the circumference direction. A combination of Doppler cooling and resolved-sideband cooling can be used to cool the ions to the ground state of the MHz trap \cite{Monroe1995,Leibfried2003}. The system is in the ground state of the ring-shaped trapping potential after ramping down the pinning potential adiabatically. For  $T^*=1.1$  nK, the ramping down time should be longer than 7 ms. Another way to achieve an ultralow temperature of ions is to put the ions near ultracold neutral atoms \cite{Zipkes2010}, which has been cooled to below 0.5 nK by adiabatic decompression\cite{Leanhardt2003}.

The ions need to be laser cooled to the ground state of a MHz trap before they can be adiabatically transferred to the ground state of the ring trap. The ground state cooling of a MHz trap can be achieved with resolved sideband cooling \cite{Monroe1995,Leibfried2003}. Recently, a ground state
population of $99\pm 1 \%$ has been achieved for a single ion oscillating at 585 kHz in a room temperature
trap with a sub-Hertz heating rate, corresponding to a temperature of only 6~$\mu$K \cite{poulsen2012}.

The electric field noise on the trap electrodes should be reduced to minimize heating. The fundamental  noise is the Johnson noise in the trap circuitry. However, real experiments are dominated by anomalous heating that is related to the contaminations of the surface of electrodes. By cleaning the surface of electrodes in situ with an argon-ion beam, Hite \emph{et al} \cite{hite2011} reduced the heating rate by two orders of magnitude at room temperature. The heating rate can also be reduced dramatically by reducing the temperature \cite{Labaziewicz2008}. The static stray electric fields should also be minimized. Recently, the stray electric field has been compensated to about 0.1 V/m  for optical trapping of an ion in a linear trap \cite{Schneider2010}. This value can be reduced further with a better design of the ion trap, for example, using only concentric planar ring electrodes \cite{clark2012}. Such a planar ring multiple trap  needs only RF voltages and is symmetric in rotation.

In conclusion, we propose a method to create and observe a space-time crystal experimentally with trapped ions. We also discuss about how to create a time quasicrystal. The space-time crystals of trapped ions provide a new dimension for studying many-body physics, and may have potential applications in
 quantum information for simulating other novel states of matter.

We thank H. H{\"{a}}ffner and C. Monroe for helpful discussions. This work was funded by Ernest S. Kuh Endowed Chair Professorship and Miller Professorship at UC Berkeley, and partially funded by NSF Nanoscale Science and Engineering Center (CMMI-0751621). Z.X.G and L.M.D were supported by NBRPC (973 Program) 2011CBA00300 (2011CBA00302), the IARPA MUSIQC program, the ARO and the AFOSR MURI program. Z.Q.Y. was supported by NBRPC (973 Program) 2011CBA00300 (2011CBA00302), NNSFC 61073174, 61033001, 61061130540, 11105136, and Postdoc Research Funding of China Grant 20110490829. H.T.Q was supported by NSF Grant No. DMR-0906601.

Note added:  After the completion of this project, we become aware of several related works \cite{Chernodub2012,Zhao2012,shapere2012b}.

\bibliography{spacetime}

\end{document}